\documentclass[aps,prl,twocolumn,superscriptaddress,floatfix,longbibliography]{revtex4-1}

\usepackage{amssymb}
\usepackage{graphicx}
\usepackage{dcolumn}
\usepackage{bm}
\usepackage{amsmath}
\usepackage{ulem}
\usepackage[colorlinks,linkcolor=magenta,citecolor=blue,urlcolor=blue]{hyperref}

\begin{document}

\title{Hierarchy of localized many-body bound states in an interacting open lattice}
\author{Yanxia Liu}
\email{yxliu-china@ynu.edu.cn}
\affiliation{School of Physics and Astronomy, Yunnan Key Laboratory for Quantum Information, Yunnan University, Kunming 650091, PR China}

\author{Shu Chen}
\email{schen@iphy.ac.cn}
\affiliation{Beijing National Laboratory for Condensed Matter Physics, Institute of
Physics, Chinese Academy of Sciences, Beijing 100190, China}
\affiliation{School of Physical Sciences, University of Chinese Academy of Sciences,
Beijing, 100049, China}

\begin{abstract}

We unveil the mechanism for the formation of puzzled boundary-localized bound states in a spinless fermionic open lattice with nearest-neighbor interactions. By solving the Bethe-ansatz equation analytically, we uncover asymmetrical string solutions corresponding to the boundary-localized bound states, which emerge in systems with at least three particles.  
The localized bound states can become bound states in continuum in a suitable parameter region.
When the number of particles increases to five or more, additional bound states away from the edge are also observed. Through rigorous analysis, we derive recurrence relations of the quasi-momentum of the localized states as a function of the number of particles, predicting the presence of hierarchy of localized many-body bound states in interacting open lattices.

\end{abstract}

\maketitle


{\it Introduction.-} Exactly solvable quantum lattice models, e.g. Fermi-Hubbard model \cite{Lieb68}, Heisenberg spin chain \cite{Bethe31,Faddeev79}, and Kondo model \cite{Andrei80,Wiegmann80}, provide crucial insights for understanding both the ground state and thermodynamical properties of strongly correlated systems \cite{Takahashi99,XWGuan13,WYCS}. In principle, the spectra can be obtained by solving Bethe-ansatz (BA) equations, and the construction of thermodynamics relies on the string hypothesis of the Bethe roots \cite{Takahashi99,Takahashi72,Takahashi73}. A n-string solution is composed of n rapidities with the same real part but different imaginary parts,  which distribute symmetrically to the real axis.
The string solutions describe bound states of constituent particles in a variety of integrable many-body systems \cite{Bethe31,McGuire,HaoYJ,Lieb,Caux,WuCJ}.
Bethe strings have recently been observed
in quantum magnets \cite{Wu_Loidl, Wang_Lake} and quantum gases \cite{Horvath}.  Recent studies also show that
Bethe strings are relevant to anomalous transport properties of integrable spin chains and the emergence
of the Kardar-Parisi-Zhang universal classes \cite{KPZ86, Ljubotina,Gopalakrishnan,Ilievski,Jepsen,Scheie,Wei,Rosenberg}.
However it is still not clear whether the BA solutions with string hypothesis fully span the whole Hilbert space for a given integrable model.


Although the change of boundary condition usually does not alter the thermodynamical properties of a quantum many-body system, the system may support some localized boundary bound states under open boundary condition (OBC), which is absent under periodic boundary condition (PBC).
It is well known that eigenstates of an open chain are composed of standing waves.
Generally OBC does not give rise to localized quantum states for a topologically trivial non-interacting system, unless boundary fields or impurities are applied. Recently, numerical results uncovered the existence of localized edge states in three-particle systems of the spinless fermion model with nearest-neighbor interaction \cite{Pinto09} and Bose-Hubbard model under OBC \cite{Pinto09, Sun}. This observation is somewhat counterintuitive, as the localized state does not appear in the single-particle and two-particle systems, suggesting that the emergence of localized edge states in this system is due to the interplay of boundary effect and many-body interaction. The multiparticle boundary states can become bound states in continuum (BICs) with suitably chosen system parameters, where their energy levels enter the continuous spectrum but the states retain spatial localization \cite{Sun}. So far a theoretical understanding of the mechanism for the emergence of localized states in the interacting open chain is still lack.
It is also unclear whether localized bound states emerge for all the many-particle systems with the particle number larger than three?

In this Letter, we utilize the exact solution of  an interacting spinless fermions in one-dimensional lattices and scrutinize the solutions of BA equation under OBC, aiming to unveil the formation mechanism of localized many-body bound states analytically. As usual, the symmetric string solutions of the BA equation correspond to the bulk bound states. Surprisingly, we find the existence of asymmetric string solution,  corresponding to the localized bound states.
These asymmetric solutions do not exit under PBC and  escape from the eyes of the predecessors. The wave functions and the eigenvalues of the many-body localized edge states have a simple analytical form. We find that the localized bound states with different fermion numbers are governed by some recurrence relations, with which a rule to discriminate the bound states with multi-fermions is obtained. With these analytical results, we can determine the conditions for the formation of localized many-body bound states. Our study unveils analytically the existence of hierarchy of localized many-body bound states in the interacting open lattice, which provides a firm ground for understanding the novel boundary phenomenon induced by the interplay of boundary effect and many-body interaction.

\begin{figure*}[tbp]
\includegraphics[width=0.96\textwidth]{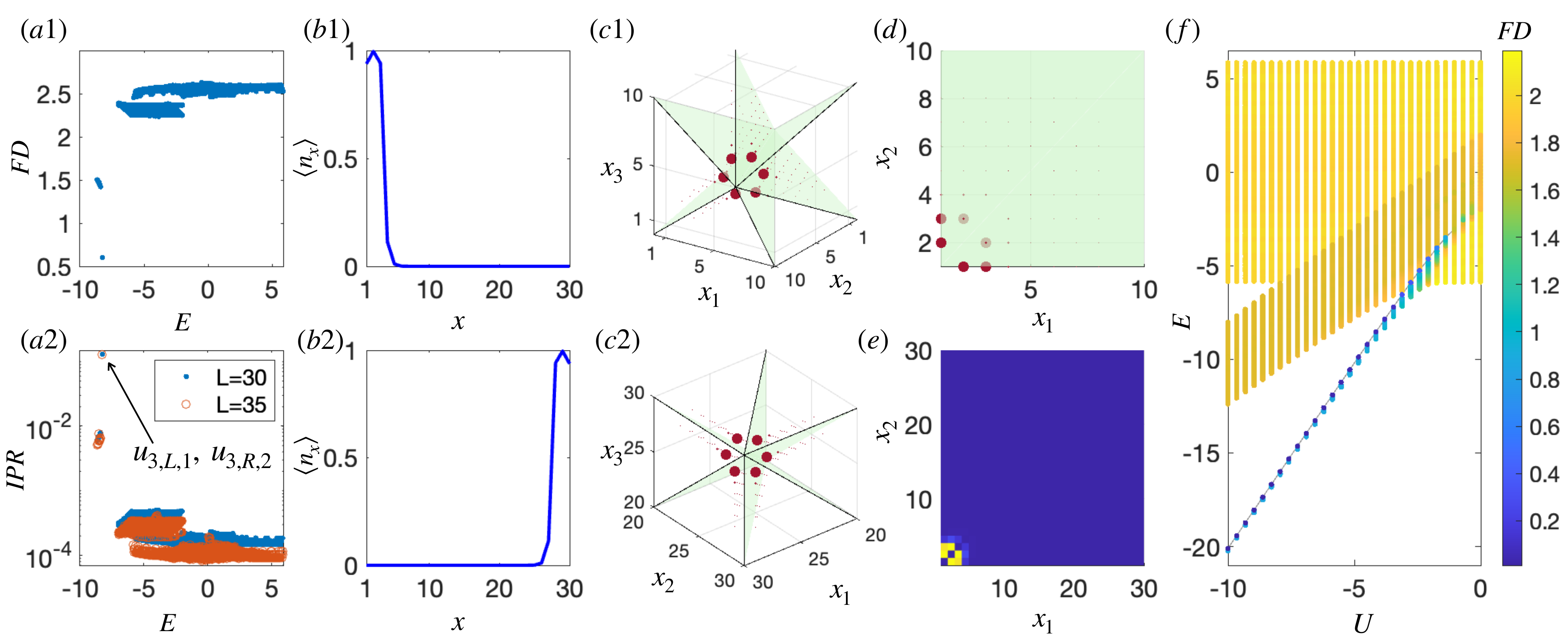}
\caption{(a)FDs and IPR for eigenstates with different eigenvalues for the system with $U=-4$, and $N=3$. The length of the lattice is chosen to be $L=30$ (blue dots) and $L=35$ (red circles). (b) Site occupancies for two localized bound states. (c) Wave function density distribution of two two localized bound states. The size of the points represents the density at that position. (d) Top view of (c1) along the $x_3$-axis. (e) The reduced density distribution $n(x_1,x_2)=\int dx_3 | \psi(x_1,x_2,x_3)|^2 $ for the left localized bound states. (f) FD of different eigenstates as a function of corresponding eigenvalue $E$ and the interaction $U$ with $N=3$ and $L=30$, respectively. The gray lines represent $E_{lb3,1}=2U+1/U$.}
\label{fig1}
\end{figure*}

{\it Model and solutions.-}
We consider a one-dimensional spinless fermion model with nearest-neighbour interaction under open boundary condition:
 \begin{align}\label{Hamiltonian}
 \hat{H}=  \sum_{x=1} ^{L-1}  \left [-t \hat{c}_x^{\dag}  \hat{c}_{x+1} -t \hat{c}_{x+1} ^{\dag}  \hat{c}_x +U \hat{n}_x  \hat{n}_{x+1} \right],
\end{align}
where $\hat{c}_x^{\dag}$, $\hat{c}_{x}$ and $ \hat{n}_x$ are fermion creation, annihilation and occupation number operators at site $x$,  $t$ is hopping amplitudes, and $U$ is the interaction strength between two particles at nearest neighbor sites and we shall consider the case of attractive interaction $U<0$. Note that our results apply to the case of repulsive interaction ($U>0$) as well.  We set $t=1$ as the unit of energy in the sequel. Our model is equivalent to the exactly solvable XXZ Heisenberg spin chain by applying a Jordan-Wigner transformation \cite{Yang661,Yang662,Yang663,Takahashi77,Takahashi99,Jordan28}, and our results can be generalized to the spin chain directly.


The model can be exactly solved by the BA method. With wave function  $|\psi \rangle=\sum_{x_1< x_2<\cdots }u(x_1,\cdots,   x_N) \hat{c}_{x_1}^{\dag}\cdots\hat{c}_{x_N}^{\dag}| 0 \rangle$, the eigenvalue equation $H|\psi \rangle =E|\psi \rangle$ can be rewritten as homogeneous linear difference equation of $u(\cdots x_j,\cdots)$:
 \begin{align}\label{Hami}
&-\sum_{j=1}^{N} \sum_{a=\pm 1} t u(\cdots x_j+a,\cdots)+U\sum_{l<j} \delta_{x_l+1,x_j} u(\cdots x_j,\cdots)    \notag\\
&= E u(\cdots x_j,\cdots),
\end{align}
where $k_j$ are quasimomentum, $P=(\cdots,P_{j},P_{j+1} ,\cdots )$ is arrangement of $(1,2, \cdots , N)$ and $\epsilon_{P}=\{\cdots,\epsilon_{P_j},\epsilon_{P_{j+1}},\cdots \} $ with $\epsilon_{P_j}=\pm 1$. $A(\epsilon_{P}P)$ are coefficients.
The wavefunction in other regions can be obtained by using exchange antisymmetry of fermions $u(\cdots,x_k,x_{j},\cdots)=-u(\cdots,x_j,x_k,\cdots)$. The corresponding eigenvalue is given by $E=-2\sum_j  \cos k_j$, where the solution of quasimomentum $k_j$ can be obtained by solving the BA equations \cite{SM,Alcaraz}:
   \begin{align} \label{BAE}
   &e^{-i2k_{j}(L+1)} \\
   =&\prod_{i=1(\neq j)}^{N} \frac{1+e^{i(k_{j}+ k_{i})}+ U e^{i k_{i} }}{1+e^{i( k_{j}+  k_{i})}+ U e^{i k_{j} }}  \frac{1+e^{i(k_{i}- k_{j})}+ U e^{-i k_{j} }}{1+e^{i( k_{i}-  k_{j})}+ U e^{i k_{i} }} . \notag
   \end{align}
Here $j=1,\cdots,N$ and $N$ is the particle number.  For convenience, we define $\gamma_j=e^{ik_j}$ and the rapidity $\lambda_i$ is given by $\lambda_i=\tanh^{-1}\left[\tan \frac{k_i}{2}/\tan \frac{\beta}{2}\right]$ with $\beta=\cos^{-1} \frac{|U|}{2}$  when $|U|<2$, and $\lambda_i=\tan^{-1}\left[\cot \frac{k_i}{2}/\coth \frac{\beta}{2}\right]$ with $\beta=\cosh^{-1} \frac{|U|}{2}$, when $|U|>2$.

For the two-fermion system ($N=2$), the spectrum of the system is composed of two continuous bands, corresponding to the two-particle scattering state and the two-particle bound state. For the scattering states, $E=-2 \cos k_1-2 \cos k_2$ with the quasimomentum $k_{1},k_{2}\in[0,2\pi]$. For the bound states, $\gamma_1=-U/(1+e^{-iQ})$ and $\gamma_2=-(1+e^{iQ})/U$, $E_{2b,2}=U +\frac{4\cos^2 \frac{Q}{2}}{U}$, where $Q=k_1+k_2\in [0,4\pi]$ is the center of mass momentum. When $|U|>2$, the solutions can convert into the familiar string hypothesis form: $\lambda^{(2)}_1=\lambda^{(2)}+i\frac{\beta}{2}$ and $\lambda^{(2)}_2=\lambda^{(2)}-i\frac{\beta}{2}$, where $\lambda^{(2)}\in\left[0,~2\pi\right]$ is related to both $U$ and $K$. $\lambda^{(2)}_1$ and $\lambda^{(2)}_2$ are called as $2$-string solutions, which indicates that two particles are bound together. Apart from these two types of solutions, the system has no other solutions \cite{SM}.


Now we consider the three-fermion case. When $L\rightarrow\infty$, the solutions of the states of the three asymptotically free fermions are $k_1,~k_2,~k_3\in \left[ 0,2\pi \right)$, and the corresponding energy $E \in \left[-6,~6\right]$; The energy of scattering states of a dimer and a fermion is $E_{2b,3}=U+4\cos^2\frac{Q}{2}/U-2\cos k_3$, with $Q/2,~k_3\in \left[ 0,2\pi \right)$. The corresponding rapidities are composed of a 2-string solution  $\lambda^{(3)}_1=\lambda^{(2)}_1$, $\lambda^{(3)}_2=\lambda^{(2)}_2$ and a real $\lambda^{(3)}_3$;
Solutions of trimer states are given by
 \begin{equation}\label{gamma3}
\gamma_1=\frac{e^{iK}(U^2-1)}{1-e^{iK}U},~\gamma_2=\frac{1-e^{iK}U}{e^{iK}-U},~\gamma_3=\frac{e^{iK}-U}{U^2-1},
\end{equation}
where $K=k_1+k_2+k_3$ is the total quasimomentum.
The spectrum of trimer states is given by $E_{3b,3}=2(U^3-\cos K)/(U^2-1)$ with $K \in \left[ 0,2\pi \right)$. The corresponding rapidity solutions are $\lambda^{(3)}_1=\lambda^{(3)}+i\beta$, $ \lambda^{(3)}_2=\lambda^{(3)}$, $ \lambda^{(3)}_3=\lambda^{(3)}-i\beta$ with $\lambda^{(3)}\in\left[0,~2\pi\right]$, which are so-called $3$-string solution. The 3-string solutions suggest that the trimer states can be viewed as a bound state composed of three particles with the bound energy determined by the imaginary part $\beta$.

While the total quasimomentum $K$ is always real for the string solutions, we note that  $K$ can have complex solutions under OBC. Solving Eqs. \eqref{BAE} in the limit of $L \rightarrow \infty$, we can get $e^{iK}=U$ or $1/U$. When $e^{iK}=U$ and $e^{iK}=1/U$, two equivalent solutions will be obtained. Without loss of generality, we let $e^{iK}=U$. Then we get
\begin{equation}
\gamma_1=-U, ~~ \gamma_2= \infty, ~~ \gamma_3= 0.  \label{gamma}
\end{equation}
This solution looks strange and has been ignored as a meaningless solution. Here we treat this solution seriously and unveil its physical meanings. The corresponding rapidity solutions can be written as
\begin{align} \label{lambda3}
\lambda^{(3)}_1=i\frac{3}{2}\beta,~ \lambda^{(3)}_2=i\frac{1}{2}\beta,~ \lambda^{(3)}_3=-i\frac{1}{2}\beta,
\end{align}
which distribute asymmetrically on the imaginary axis. Note that for the localized bound state, $\lambda_i=\tanh^{-1}\left[\cot \frac{k_i}{2}/\cot \frac{\beta}{2}\right]$ are used, when $|U|<2$.      Normally, the string solutions of bound states are symmetric about the real axis of $\lambda_i$, meaning $\lambda_i$ either appears in complex conjugate pairs or as real numbers. However, the solutions of (\ref{lambda3}) appear as complex numbers without conjugate pairs. These solutions are specific to the OBC and describe the many-body bound states, which localized on the boundaries. To distinguish with the usual string solutions, we refer them as the boundary string solutions \cite{Skorik}.  The corresponding wavefunctions are analytically given by
\begin{align}
u_{3,L,1}&= e^{-ik_3x_1-ik_2x_2-ik_1x_3}-e^{ik_2x_1+ik_3x_2-ik_1x_3}\notag\\
&\propto \delta_{x_1,x_2-1}e^{-ik_1x_3},\label{u31}\\
\text{and}~~ ~ ~  u_{3,R,2}&\propto e^{ik_1x_1}\delta_{x_2,x_3-1},\label{u32}
\end{align}
which are defined in the region $x_1<x_2<x_3$.

For details of the calculation, see the Supplementary Material \cite{SM}. From Eq. \eqref{u31} and \eqref{u32}, we can see that when $U<-1$ fermions are only distributed in the plane $x_1=x_2-1$, and the density profile decays exponentially along the positive direction of $x_3$.

To testify our analytical results and gain an intuitive understanding, we employ the exact diagonalization method to obtain the eigenvalues and eigenstates, with which we plot the fractal dimension ($\text{FD}$) of the corresponding eigenstates as a function of eigenenergies in Fig. \ref{fig1} (a1) to characterize the different continuum band. Here the fractal dimension of a finite size system is defined as
 \begin{equation}\label{FD}
\text{FD}_j (L) =- \ln (\text{IPR}_j)/\ln(L),
\end{equation}
where $\text{IPR}_j =\sum_{x_1,\cdots, x_N}|u_j(x_1,\cdots, x_N)|^4$ is the inverse participation ratio. The FD is obtained via $\text{FD}_j=\lim_{L\rightarrow \infty}\text{FD}_j (L)$. The FD goes to $0$ for the bound states, $1$ for the trimer states with $E=E_{3b,3}$, $2$ for the states with $E=E_{2b,3}$, and  $3$ for the scattering states with $E=E_{3s}$. The continuum bands from top to bottom in Fig. \ref{fig1} (a) are $E_{3s}$, $E_{2b,3}$, and $E_{3b,3}$, respectively. At the bottom, there are two overlapping points, of which the $FD \approx 0.5$. The corresponding states are bound states. Note that FD of localized states goes to $0$ since their IPR are not sensitive to the lattice size. To validate this, we plot the IPR of the corresponding eigenstates as a function of eigenenergies in Fig. \ref{fig1} ((a2) with different lattice sizes.  It is shown that the IPRs of states in continuum bands change with the lattice size, whereas IPRs of the localized states are not sensitive to the change in lattice size. Fig. \ref{fig1}(b1) and (b2) display density profiles $\langle n_x \rangle = \langle \psi  |\hat{c}_x^{\dag}  \hat{c}_{x}|\psi \rangle $ with $x=1,2,\cdots L$ for two degenerate bound states, indicating that the bound states are localized at the left and right edges of the lattice.


In Fig. \ref{fig1}(c) and (d), we present numerical results of the density distribution for the localized bound states, which are consistent with the analytical results.  In this state $|p|=|k_2-k_3|=\infty$, the position $x_2-x_1$ can be measured exactly, i.e. $x_2-x_1=1$, from the uncertainty principle. According to the exchange antisymmetry, we can know the behavior of wavefunction in other regions. The wave function indeed only distributes over certain fixed planes: $x_i-x_j=\pm 1$ with $i\neq j =1,~2,~3$. The energy of this states is $E_{lb,3,1}=\frac{1}{U}+2U$, which are in good agreement with the numerical results, as shown in  Fig. \ref{fig1} (f).  From Fig. \ref{fig1} (f)  it can be seen that the energies of the localized bound states fall into the continuous spectrum, when $-3<U<-1$. In other word, this system has BICs \cite{Sugimoto23,ZhangJM12,ZhangJM13,Huang23,Liu24}.

\begin{figure}[tbp]
\includegraphics[width=0.48\textwidth]{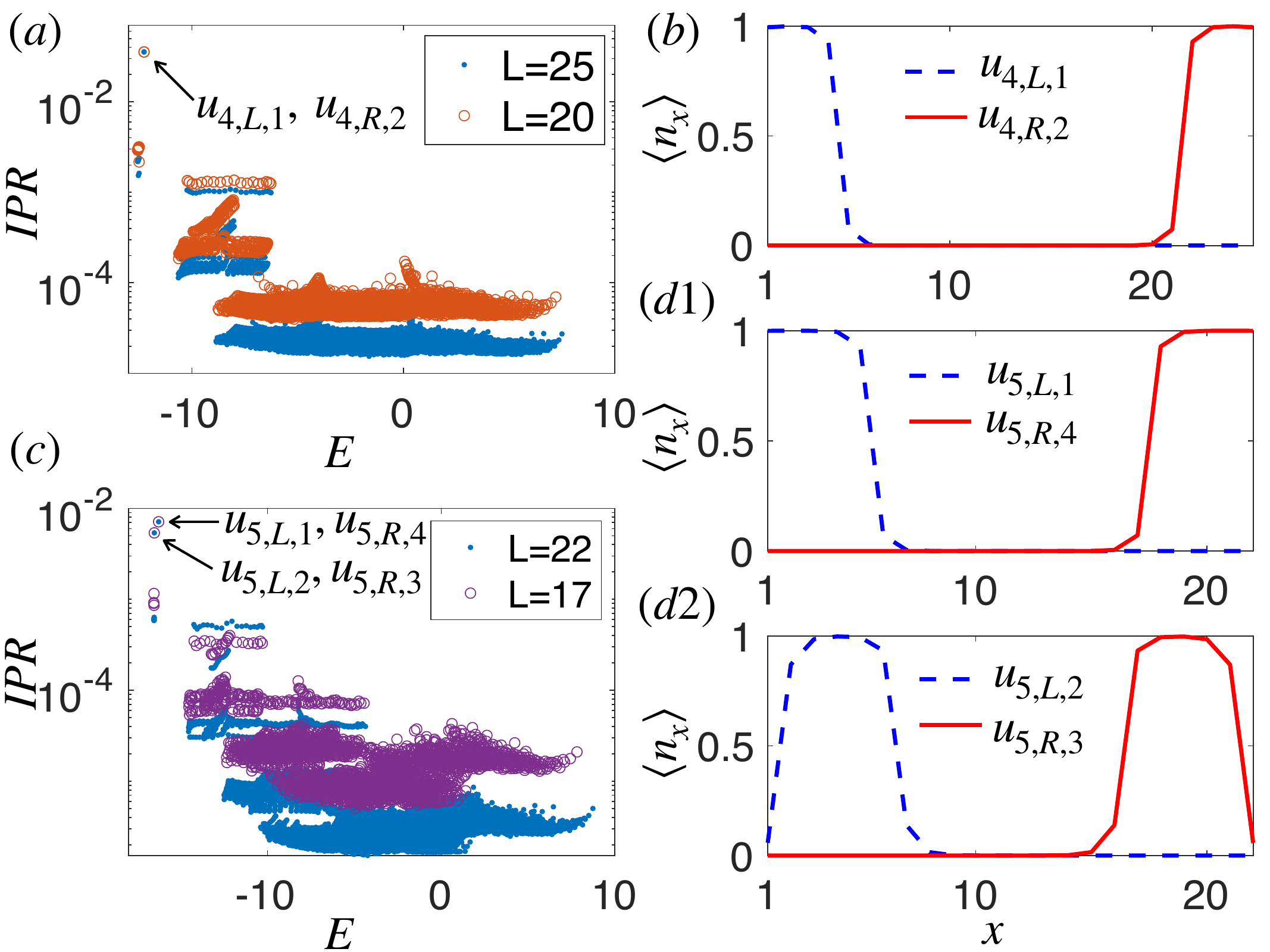}
\caption{(a) IPR for eigenstates with different eigenvalues for the system with $U=-4$, $N=4$. The length of the lattice is chosen to be $L=25$ (blue dots) and $L=20$ (red circles). The IPR of the localized bound state with different $L$ are the same. (b) The expectation values of occupation number at each site $n_x$ of the localized bound states $u_{4,L,1}$ and $u_{4,R,2}$ in (a). (c) IPR for eigenstates with different eigenvalues for the system with $U=-4$, $N=5$. The length of the lattice is chosen to be $L=22$ (blue dots) and $L=20$ (red circles). (d)  The expectation values of occupation number at each site $n_x$ of the localized bound states $u_{5,L,i}$ and $u_{5,R,i}$ in (c).}
\label{fig2}
\end{figure}



{\it Hierarchy of localized many-body bound states.-}
Next we consider cases with $N>3$ and unveil the recurrence relations for deriving hierarchy of localized many-body bound states. For $N=4$, there are six continuum bands and two localized edge states in presence.  Details can be found in the Supplementary Material \cite{SM}. Here we focus on localized bound states. 
To solve Eqs. \eqref{BAE} with $N=4$, we get the solutions for the localized bound states: $\gamma_1$, $\gamma_2$, and $\gamma_3$ satisfy Eq. \eqref{gamma} and $\gamma_{4}=g(\gamma_{1})$, where $g(x)=-1/x-U$. There are two degenerate localized bound states, as shown in Fig \ref{fig2} (a).  The corresponding wave functions are
\begin{align}
u_{4,L,1}=\delta_{x_1,x_2-1}\frac{1}{\gamma_1^{x_3} \gamma_4^{x_4} };\label{wavefunction41} \\
u_{4,R,3}=\gamma_4^{x_1}\gamma_1^{x_2}\delta_{x_3,x_4-1}.\label{wavefunction42}
\end{align}
which are two degenerate states with energy $E_{lb,4,1}=E_{lb,3,1}-1/\gamma_4-\gamma_4$. As shown in Fig. \ref{fig2} (b), the two localized bound states are localized at the left and right edge of the lattice. The conditions for $u_{4,L,1}$ and $u_{4,R,3}$ being localized edge states are: $U< (-\sqrt{5}-1)/2$. The corresponding rapidities are $\lambda^{(4)}_i=\lambda^{(3)}_i$ with $i=1,~2,~3$ and
$
\lambda^{(4)}_4=i\frac{5}{2}\beta
$,
which distribute asymmetrically on the imaginary axis.


For the case of $5$ particles, there are $4$ localized bound states, and each pair is degenerate, as shown in Fig. \ref{fig2} (c). The  solutions for these localized bound states are $\gamma_{5}=\gamma_{5,1}=g(\gamma_{4})$ and $\gamma_{5}=\gamma_{5,2}=f(\gamma_{4})=1/\gamma_1$ where $f(x)=-x-U$. The corresponding rapidities are $\lambda^{(5)}_i=\lambda^{(4)}_i$ with $i=1,~\cdots,~4$, $\lambda^{(5)}_{5}=i\frac{7}{2}\beta$ or $-i\frac{3}{2}\beta$ \cite{SM}. When $\lambda^{(5)}_{5}=i\frac{7}{2}\beta$, the wave functions are
\begin{align}
u_{5,L,1}=\delta_{x_1,x_2-1}\frac{1}{\gamma_1^{x_3} \gamma_4^{x_4}\gamma_5^{x_5} };\label{wavefunction51} \\
u_{5,R,4}=\gamma_5^{x_1}\gamma_4^{x_2}\gamma_1^{x_3}\delta_{x_4,x_5-1},\label{wavefunction52}
\end{align}
As shown in Fig. \ref{fig2} (d1), these states are localized at the left and right edge, respectively. The conditions $u_{5,L,1}$ and $u_{5,R,4}$ being localized edge states are: $U<-1.80194$ or $-1<U<-0.618$.
When $\lambda^{(5)}_{5}=-i\frac{3}{2}\beta$,
\begin{align}
u_{5,L,2}=\gamma_1^{x_1} \delta_{x_2,x_3-1}\frac{1}{\gamma_1^{x_4} \gamma_4^{x_5} };\label{wavefunction41} \\
u_{5,R,3}=\gamma_4^{x_1}\gamma_1^{x_2}\delta_{x_3,x_4-1}\frac{1}{\gamma_1^{x_5}}.\label{wavefunction42}
\end{align}
As shown in Fig. \ref{fig2} (d2), these states are localized away from the left and right edge of the lattice chain. The conditions $u_{5,L,2}$ and $u_{5,R,3}$ being localized edge states are: $U< (-\sqrt{5}-1)/2$.



 \begin{figure}[tbp]
\includegraphics[width=0.49\textwidth]{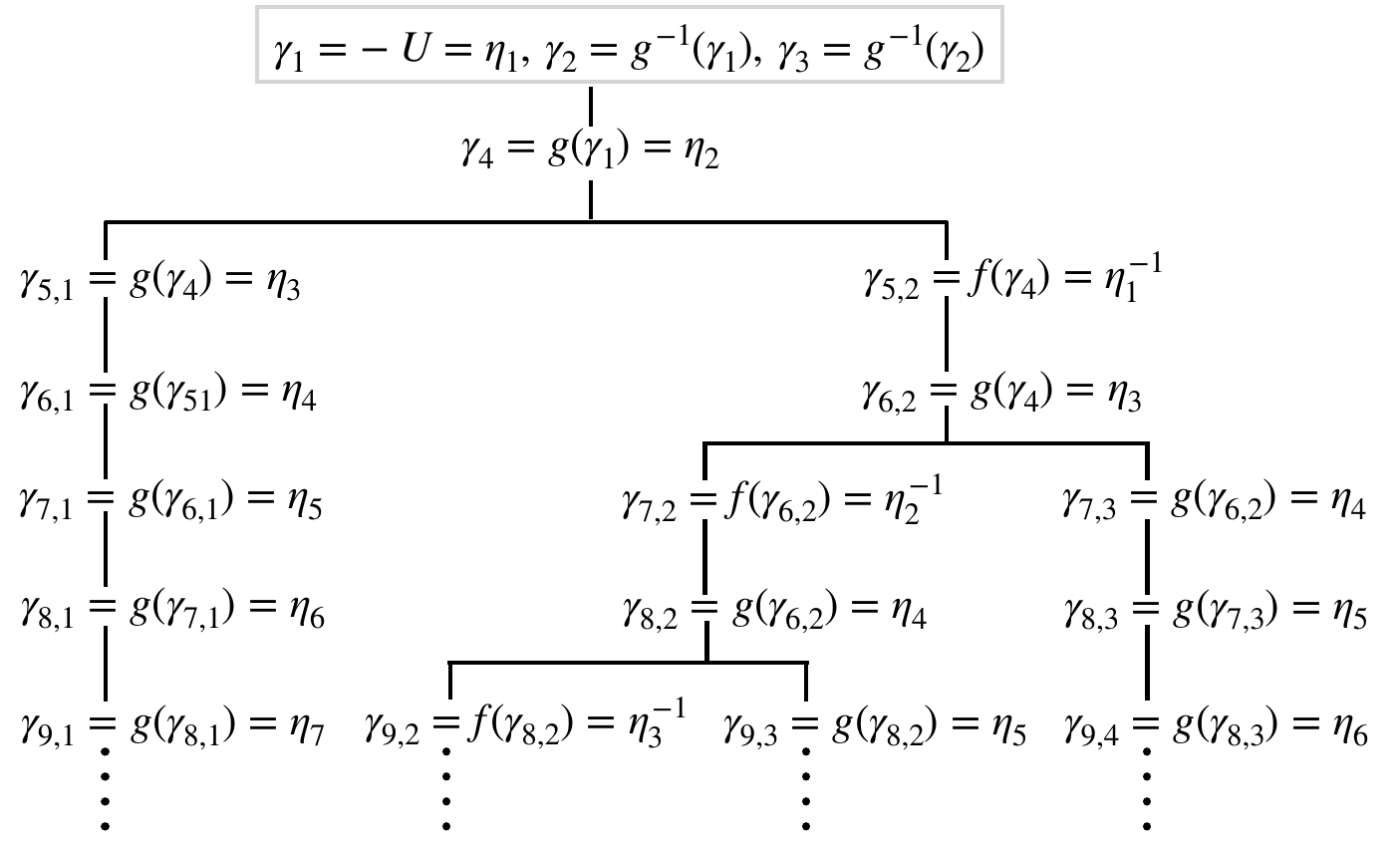}
\caption{The recurrences of $\gamma_j$ for the localized bound states. Here $g(x)=-1/x-U$ and $f(x)=-x-U$. We have used $f(g(x))=1/x$, and $g^{-1}(x)=1/f(x)$ is the inverse of $g(x)$.}
\label{fig3}
\end{figure}

In general cases, we can add fermions one by one. We discover the solutions of $\gamma_j$ for these localized many-body states fulfilling some recurrence relations, as shown in Fig. \ref{fig3}. By these recurrence relations, we see that there are $2\times[(N-1)/2] $ localized $N$-body bound states composed of $N$-fermion. With a set of $\gamma_j$, we can get the corresponding wave function:
 \begin{align}\label{wavefunctionNL}
u_{N,L,i}  =\cdots\eta_2^{x_{i-2}} \eta_1^{x_{i-1}}\delta_{x_i,x_{i+1}-1} \frac{1}{\eta_1^{x_{i+2}}\eta_2^{x_{i+3}} \cdots },
\end{align}
with $0<i\leqslant [(N-1)/2]$, and
  \begin{align}\label{wavefunctionNR}
u_{N,R,i}  =\cdots\eta_2^{x_{i-2}} \eta_1^{x_{i-1}}\delta_{x_i,x_{i+1}-1} \frac{1}{\eta_1^{x_{i+2}}\eta_2^{x_{i+3}} \cdots },
\end{align}
with $[N/2]<i< N$, where $\eta_{j+1}=g(\eta_j)$ with $\eta_1=\gamma_1=-U$, and $j>0$ is defined.
Here, the last index in the subscript of $u_{N,L(R),i}$ shows that the $i$-th particle and the $(i+1)$-th particle in the state are always positioned on neighbor sites.
The corresponding energy are $E_{lb,N,i}  =-\sum_{l=1}^{i-1} \left(\frac{1}{\eta_l}+\eta_l\right)-\sum^{N-i-1}_{l=1} \left(\frac{1}{\eta_l}+\eta_l\right)$.
It is clear that $u_{N,L,j}$ and $u_{N,R,N-j}$ are degenerate.
The conditions for $u_{N,i}$ being a localized bound state are
$
 \{|\eta_1 \cdots   \eta_{N-1-i}|>1,~ |\eta_2 \cdots   \eta_{N-1-i}|>1, ~\cdots ,~|  \eta_{N-1-i}|>1 \} ,
$
 and
$
 \{|\eta_1 \eta_2 \cdots   \eta_{i-1}|>1,~ |\eta_2 \cdots   \eta_{i-1}|>1, ~\cdots ,~|  \eta_{i-1}|>1 \}  .
$
From Eqs.\eqref{wavefunctionNL} and \eqref{wavefunctionNR} it is quite straightforward to see as long as the number of the $\eta$s is imbalanced on the different side of the delta symbol, we have a localized state on the side with fewer $\eta$s.

 The above solutions all describe situations where all particles are bound together locally, either entirely on the left side of the lattice or on the right side. The formation of localized bound states here requires at least $3$ particles. If we now have $6$ particles, is it possible for $3$ particles to be localized on the left side and the other $3$ particles to be localized on the right side? The answer is yes. Since in the limit of $L \rightarrow \infty$, the wave functions of localized many-body bound states on the left side and those on the right side have no overlap, and can be treated independently. Consequently, if the number of particles $N > 6$,  localized many-body bound states can exhibit diverse forms via the combination of left and right localized bound states.

 From the recurrence relation $\eta_{j}$, $\eta_{j+1}=-1/\eta_{j}-U$, if $U<-2$, this sequence converges to $\eta_{\infty}+1/\eta_{\infty}=-U$. And for a localized bound state with $j$ fermions under given $U$, we have $|\eta_{j-1-i}|>1$ and $|\eta_{i-1}|>1$. The recurrence relation then gives $|\eta_{j-i}|>1$ or $|\eta_{i}|>1$. Consequently, we demonstrate that $u_{j+1,L,i}$ or $u_{j+1,R,i}$ meets the  localized bound-state requirement
$
 \{|\eta_1 \cdots   \eta_{j-i}|>1,~ |\eta_2 \cdots   \eta_{j-i}|>1, ~\cdots ,~|  \eta_{j-i}|>1 \} ,
$
or
$
 \{|\eta_1 \eta_2 \cdots   \eta_{i}|>1,~ |\eta_2 \cdots   \eta_{i}|>1, ~\cdots ,~|  \eta_{i}|>1 \}.
$
So, the localized bound states are always present for $U<-2$ and $N\geqslant 3$.


{\it Conclusion.-} In conclusion, we unveil the emergence of hierarchy of localized many-body bound states in an interacting open lattice by identifying the existence of corresponding asymmetric boundary string solutions of BA equations. Our analytical results unravel the mechanism for the formation of localized many-body bound states and many-body BICs, driven by the interplay of boundary effects and interactions, providing a theoretical foundation for understanding the puzzle boundary phenomenon in quantum many-body systems.

\begin{acknowledgments}
We would like to thank Y.-S. Cao for the useful dis- cussions. This work was supported by the National Natural Science Foundation of China Grant under Nos. 12204406, No.12474287, and No.T2121001.
the National Key Research and Development Program of China (2021YFA1402104).
\end{acknowledgments}


%

\end{document}